\documentclass[conf, 10pt, letter]{IEEEtran}
\usepackage[utf8]{inputenc}
\usepackage{amsmath}
\usepackage{amsfonts}
\usepackage{amssymb}
\usepackage{amsthm}
\usepackage{commath}
\usepackage{graphicx}
\usepackage{bm}
\usepackage{bbm}
\usepackage{multirow}
\usepackage{epstopdf}
\usepackage{physics}
\usepackage{mathtools}
\usepackage{cases}
\usepackage{enumitem}
\usepackage{cite}
\usepackage{xcolor}
\usepackage{cases}
\usepackage{balance}
\usepackage{caption}
\usepackage[labelformat=simple]{subcaption}

\allowdisplaybreaks
\newtheorem{theorem}{Theorem}
\newtheorem{definition}{Definition}
\newtheorem{lemma}{Lemma}
\newtheorem{proposition}{Proposition}

\theoremstyle{definition}

\theoremstyle{definition}
\newtheorem{example}{Example}
\theoremstyle{definition}

\newcommand\scalemath[2]{\scalebox{#1}{\mbox{\ensuremath{\displaystyle #2}}}}

\makeatletter
\def\IEEElabelanchoreqn#1{\bgroup
\def\@currentlabel{\p@equation\theequation}\relax
\def\@currentHref{\@IEEEtheHrefequation}\label{#1}\relax
\Hy@raisedlink{\hyper@anchorstart{\@currentHref}}\relax
\Hy@raisedlink{\hyper@anchorend}\egroup}
\makeatother
\newcommand{\subnumberinglabel}[1]{\IEEEyesnumber
\IEEEyessubnumber*\IEEElabelanchoreqn{#1}}

\makeatletter
\renewcommand*\env@matrix[1][\arraystretch]{%
  \edef\arraystretch{#1}%
  \hskip -\arraycolsep
  \let\@ifnextchar\new@ifnextchar
  \array{*\c@MaxMatrixCols c}}
\makeatother

\newcommand{\argmax}{\operatorname*{argmax}}

\newcommand{\markov}{\mathrel{\multimap}\joinrel\mathrel{-}%
\joinrel\mathrel{\mkern-6mu}\joinrel\mathrel{-}}
\newcommand{\mli}[1]{\mathit{#1}}

\begin{document}

\title{An Information Bottleneck Problem \\ with R\'{e}nyi's Entropy}

\author{%
  \IEEEauthorblockN{Jian-Jia Weng, Fady Alajaji, and Tam\'as Linder}\\
  \IEEEauthorblockA{Department of Mathematics and Statistics\\
                    Queen's University\\
                    Kingston, ON K7L 3N6, Canada\\
                    Emails: jianjia.weng@queensu.ca, \{fady, linder\}@mast.queensu.ca}
  \thanks{This work was supported in part by NSERC of Canada.}            
}


\maketitle
\begin{abstract}
  This paper considers an information bottleneck problem with the objective of obtaining a most informative representation of a hidden feature subject to a R\'{e}nyi entropy complexity constraint. 
  The optimal bottleneck trade-off between relevance (measured via Shannon's mutual information) and R\'{e}nyi entropy cost is defined and an iterative algorithm for finding approximate solutions is provided. 
  We also derive an operational characterization for the optimal trade-off by demonstrating that the optimal R\'{e}nyi entropy-relevance trade-off is achievable by a simple time-sharing scalar coding scheme and that no coding scheme can provide better performance. 
  Two examples where the optimal Shannon entropy-relevance trade-off can be exactly determined are further given. 
\end{abstract}


\section{Introduction}
In the past decade, the optimization of information measures such as entropy, cross-entropy, and mutual information has been widely and successfully adopted in machine learning algorithms \cite{amjad2019, strouse2019,goldfeld2020,zaidi2020} and transmission systems \cite{zeitler2012, winkelbauer2013, kurkoski2014, meidlinger2019, nguyen2020, stark2020}. 
In particular, numerous results are related to the so-called information bottleneck (IB) method \cite{tishby1999} whose objective is to extract from observed data the maximal relevant information about a hidden variable subject to a mutual information complexity constraint. 
Significant efforts are still devoted to studying the IB method and its variants, including its variational approximation \cite{alemi2016}, its application to analyze the effectiveness of deep neural networks \cite{shwartz2017}, and its generalizations \cite{hsu2018, shahab2020}. 
This paper studies a constrained information optimization problem that is close to a deterministic variant of the IB method, the so-called deterministic IB (DIB) method \cite{strouse2017}. 

The DIB method was proposed to capture the notion of compression in the IB method. 
In view of Shannon's lossless source coding theorem \cite{shannon1948}, the authors of \cite{strouse2017} suggested replacing the mutual information complexity constraint in the IB method with a Shannon entropy complexity constraint to take into account the cost of storing the extracted information. Here, the storage cost is assumed to vary linearly with the length of the codeword that represents the extracted information. 
Motivated by Campbell's source coding theorem \cite{campbell1965}, an extension of Shannon's result, we consider a R\'{e}nyi entropy \cite{renyi1960} constraint to associate the codeword length with an exponential storage cost (which is more appropriate for applications where the processing cost of decoding and buffer overflow problems caused by long codewords are significant). 
As Shannon entropy is a limiting case of R\'{e}nyi entropy \cite{renyi1960} (as the order goes to $1$), this consideration extends the DIB method in a certain sense. 

Our R\'{e}nyi extension may provide a way to improve the performance of the DIB method in machine learning tasks \cite{strouse2019} or other applications such as channel quantization \cite{kurkoski2014} and relay transmission \cite{nguyen2020}. In addition, the use of R\'{e}nyi entropy generated interest in its own right in information theory and it has played an important role in a variety of studies, including generalized source-coding cut-off rates \cite{csiszar1995, chen2001}, quantization \cite{kreitmeier2011}, encoding tasks \cite{bunte2014}, guessing \cite{bracher2019}, information combining \cite{hirche2020}, generative deep networks \cite{bhatia2020}, etc. 
It is thus of interest to examine the role of R\'{e}nyi entropy in bottleneck problems.

We now formulate our bottleneck problem. Consider a pair of discrete random variables $(Y, X)$ with joint probability distribution $P_{Y, X}$ over a finite alphabet $\mathcal{Y}\times\mathcal{X}$ and another (representation) random variable $W\in\mathcal{W}$, which form a Markov chain: $Y{\markov}X{\markov}W$, i.e., $P_{Y, X, W}=P_{Y, X}P_{W|X}$. 
The objective is to determine the maximal amount of relevant information about $Y$ that can be extracted from $X$ and conveyed in $W$ subject to a R\'{e}nyi entropy constraint $H_\alpha(W)\le \gamma$ for $\alpha\in(0, 1)$,\footnote{Although $H_{\alpha}(W)$ is defined for $\alpha\in(0, 1)\cup (1,\infty)$, we only consider $\alpha\in(0, 1)$ since $H_{\alpha}(W)$ with such $\alpha$ has a clear operational meaning \cite{campbell1965}.} where $H_\alpha(\cdot)$ denotes the R\'{e}nyi entropy of order $\alpha$ \cite{renyi1960}. 
We call this problem an information-R\'{e}nyi entropy bottleneck problem 
and study the function $F_{\alpha, M}:\mathbb{R}_{\ge 0}\to\mathbb{R}_{\ge 0}$ in connection with the problem:
\begin{IEEEeqnarray}{C}
  F_{\alpha, M}(\gamma):=\max_{\substack{P_{W|X}: Y\markov X\markov W\\ |\mathcal{W}|\le M, H_{\alpha}(W)\le\gamma}} I(Y; W) \label{eq:GEIBC}
\end{IEEEeqnarray}
where $\gamma\ge 0$, $M$ is a finite positive integer, and $|\mathcal{W}|$ denotes the cardinality of $\mathcal{W}$. 
To simplify our presentation, we let $F_{1, M}$ denote \eqref{eq:GEIBC} with the constraint $H_{\alpha}(W)\le\gamma$ replaced by a Shannon entropy constraint $H(W)\le\gamma$, which is a constrained optimization formulation of the DIB problem in \cite{strouse2017}. 

Define $\bar{I}_{\alpha,M}$ as the upper concave envelope of $F_{\alpha, M}$ for $\alpha\in(0, 1]$; the function $\bar{I}_{\alpha,M}$ can be expressed by \cite[Corollary 17.1.5]{rockafellar1970}
\vspace{-0.4cm}\begin{IEEEeqnarray}{l}
  \bar{I}_{\alpha,M}(\gamma)=\max\sum_{i=1}^2\lambda_i F_{\alpha,M}(\gamma_i)
  \label{eq:barI}
\end{IEEEeqnarray}
where the maximum is taken over all convex combinations of two pairs $(\gamma_i, F_{\alpha,M}(\gamma_i))$, $i=1, 2$, such that $\sum_{i=1}^2\lambda_i \gamma_i=\gamma$. 
In this paper, we show that the function $\bar{I}_{\alpha, M}$ describes the optimal R\'{e}nyi entropy-relevance trade-off for our bottleneck problem. 
Specifically, we establish an operational characterization of $\bar{I}_{\alpha, M}$ for any ${\alpha\!\in\!(0,1)}$ and finite $M$. 
This finding is analogous to the IB coding theorem \cite{gilad2003} and clarifies the operational meaning of the information quantities in \eqref{eq:GEIBC}.
We note that a closed-form expression of $\bar{I}_{\alpha, M}$ is only available for very special cases. We also derive bounds for $\bar{I}_{\alpha, M}$ and provide numerical methods to compute $\bar{I}_{\alpha, M}$.

The rest of this paper is organized as follows. 
In Section~II, the system model is given and the IB and DIB results are reviewed.  
In Section~III, bounds and properties for $\bar{I}_{\alpha, M}$ are established, followed by the derivation of an operational characterization. 
In Section~IV, methods to compute $\bar{I}_{\alpha, M}$ and two examples are given. 
Conclusions are drawn in Section~V. 


\section{Preliminaries}
Given discrete random variables $A_i$ on a common alphabet $\mathcal{A}$, $i=1, 2, \dots, n$, we let $A^n =\allowbreak (A_1, \allowbreak A_2, \allowbreak \dots, \allowbreak A_n)$. 
Throughout this paper, we assume the following system model when developing information-theoretic results. The system input is a sequence of independent and identically distributed (i.i.d.) random variables $X_i\in\mathcal{X}$, $i=1, 2, \dots, n$, which is correlated with another sequence of i.i.d. hidden variables $Y_i\in\mathcal{Y}$. The joint probability distribution is given by $P_{Y^n, X^n}(y^n, x^n) =\allowbreak\prod_{i=1}^n P_{Y, X}(y_i, x_i)$ for some joint distribution $P_{Y, X}$. 
The goal is to transform $X^n$ into the most informative $W^n$ subject to a complexity constraint. 
In \cite{gilad2003}, a complete coding theorem is derived for the IB method, but there seems to have no corresponding result for the DIB method. 
One of our objectives is to fill this gap under a more general framework. 

We begin with the definition of R\'{e}nyi entropy \cite{renyi1960} of order $\alpha\in(0,1)\cup(1, \infty)$ of a random variable $W$ with alphabet $\mathcal{W}$ and distribution $P_{W}$: 
\vspace*{-0.2cm}\begin{IEEEeqnarray}{C}
  H_{\alpha}(W):=\frac{1}{1-\alpha}\log_2\scalemath{0.9}{\left(\sum_{w\in\mathcal{W}}P^{\alpha}_{W}(w)\right)}.\nonumber\IEEEeqnarraynumspace
\end{IEEEeqnarray}
Some properties of $H_{\alpha}(W)$ for ${\alpha\!\in\!(0,1)}$ are summarized \cite{van2014}:  
\begin{itemize}
  \item $H(W)=H_1(W):=\lim_{\alpha\to 1} H_{\alpha}(W)$; 
  \item $0\le H_{\alpha}(W)\le \log_2 |\mathcal{W}|$;
  \item $H_{\alpha}(W)$ is non-increasing in $\alpha$; 
  \item $H_{\alpha}(W)$ is concave in $P_W$ for $\alpha\in[0,1]$;
  \item $H_{\alpha}(W)$ is continuous in $\alpha$ at any ${\alpha\!\in\!(0,1)}$ and finite $\mathcal{W}$.
\end{itemize}

In this paper, all information quantities are measured in bits. 
We next review some IB and DIB results related to our work.  

\subsection{The IB Method \cite{tishby1999} and A Complete Coding Theorem \cite{gilad2003}}
In short, the IB method aims to extract from an observation $X$ the most relevant representation $W$ about a hidden variable $Y$ under a complexity constraint. For any fixed $P_{Y, X}$, such an objective is associated with the function $\bar{I}_{\text{IB}}:\mathbb{R}_{\ge 0}\to\mathbb{R}_{\ge 0}$ defined as
\begin{IEEEeqnarray}{C}
  \bar{I}_{\text{IB}}(r):=\max_{\substack{P_{W|X}: Y\markov X\markov W\\ I(X; W)\le r}} I(Y; W) \label{eq:IB}
\end{IEEEeqnarray}
which is non-decreasing, continuous, and concave in $r$ \cite{witsenhausen1975}. One can also set $|\mathcal{W}|=|\mathcal{X}|+1$ without changing $\bar{I}_{\text{IB}}(\gamma)$. 
An operational interpretation of the $\bar{I}_{\text{IB}}$ function is described next. 
\begin{definition}
  A $(2^{\mli{nr_n}}, n)$ IB code consists of an encoding function $\mathcal{E}_n: \mathcal{X}^n\to \{1, 2, \dots, 2^{\mli{nr_n}}\}$ and a decoding function $\mathcal{D}_n: \{1, 2, \dots, 2^{\mli{nr_n}}\}\to\mathcal{W}^n$. 
\end{definition}

The average symbol-wise mutual information between $Y^n$ and $W^n$ associated with the above code is computed as $\eta_n:=\frac{1}{n}\sum_{i=1}^{n} I(Y_i; W_i)$, where $W^n=\mathcal{D}_n(\mathcal{E}_n(X^n))$. 

\begin{definition}
  \label{def:IBcodes}
  A complexity-relevance pair $(r, \eta)$ is said to be achievable if there exists a sequence of IB codes $\{\mathcal{E}_n, \mathcal{D}_n\}$ such that $\limsup_{n\to\infty} r_n\le r$ and $\liminf_{n\to\infty} \eta_n \ge \eta$. 
  The achievable IB region $\mathcal{R}_{\emph{IB}}\subset \mathbb{R}_{\ge 0}^2$ is defined as the closure of all achievable pairs. 
\end{definition}

Letting $I_{\text{IB}}(r)\! =\!\max(\eta\!: (r, \eta){\in}\mathcal{R}_{\text{IB}})$, Gilad {\it et al.} proved the following proposition in \cite{gilad2003}. Here, we rephrase their original statement in \cite[Theorem 2]{gilad2003} in terms of the functions $I_{\text{IB}}$ and $\bar{I}_{\text{IB}}$, which is more convenient for our use.

\begin{proposition}[\hspace{-0.04em}\cite{gilad2003}]
  $I_{\emph{IB}}(r)=\bar{I}_{\emph{IB}}(r)$ for $r\ge 0$.\label{prop:IBCodingThm}
\end{proposition}


\subsection{The DIB Method \cite{strouse2017}}\label{sec:IIB}
As mentioned before, the DIB method borrows the source coding idea from information theory and intends to minimize the representation cost for $W$. 
Specifically, the DIB method is associated with the following optimization problem
\begin{IEEEeqnarray}{C}
  \max_{P_{W|X}: Y\markov X\markov W} \left[\beta I(Y; W)-H(W)\right] \label{eq:VF}
\end{IEEEeqnarray}
where ${\beta\in[0,\infty)}$ controls the trade-off between $I(Y; W)$ and $H(W)$. When $\mathcal{W}$ is finite, \eqref{eq:VF} is a convex optimization problem since, as one can verify, $H(W)$ is concave in $P_{W|X}$ and $I(Y; W)$ is convex in $P_{W|X}$ for any fixed $P_{Y, X}$ \cite{geiger2016}. Moreover, the feasible set of all valid conditional probability distributions $P_{W|X}$ is compact and convex. 
Therefore, we know that the maximum value of the objective function is attainable by some extreme point of the feasible set \cite[Corollary 32.3.1]{rockafellar1970}. 
In our case, the extreme points are conditional distributions $P_{W|X}$ where  $P_{W|X}(w|x)$ is either $0$ or $1$ for any $x\in\mathcal{X}$ and $w\in\mathcal{W}$, i.e., the optimizer of \eqref{eq:VF} is deterministic.\footnote{\cite{strouse2017} adopted another more complex approach to solve \eqref{eq:VF} and concluded the same deterministic structure for the maximizing distribution.}

We remark that the objective function to be maximized in \eqref{eq:VF} can be viewed as the Lagrangian corresponding to the constrained optimization problem in \eqref{eq:GEIBC} with $H_{\alpha}(W)$ replaced by $H(W)$, where $\beta$ denotes the Lagrange multiplier. 
From this viewpoint, the DIB problem is a special case of our information-R\'{e}nyi entropy bottleneck problem. 
In \cite{strouse2017}, the DIB method was empirically shown to attain a relevance level similar to that of the IB method with smaller entropy $H(W)$. 
However, unlike the IB method, the DIB method does not have an operational meaning in terms of a coding theorem.  

\begin{figure*}[!h]
  \centering
    \begin{subfigure}[b]{0.3\textwidth}
      \centering
        \includegraphics[draft=false, scale=0.435]{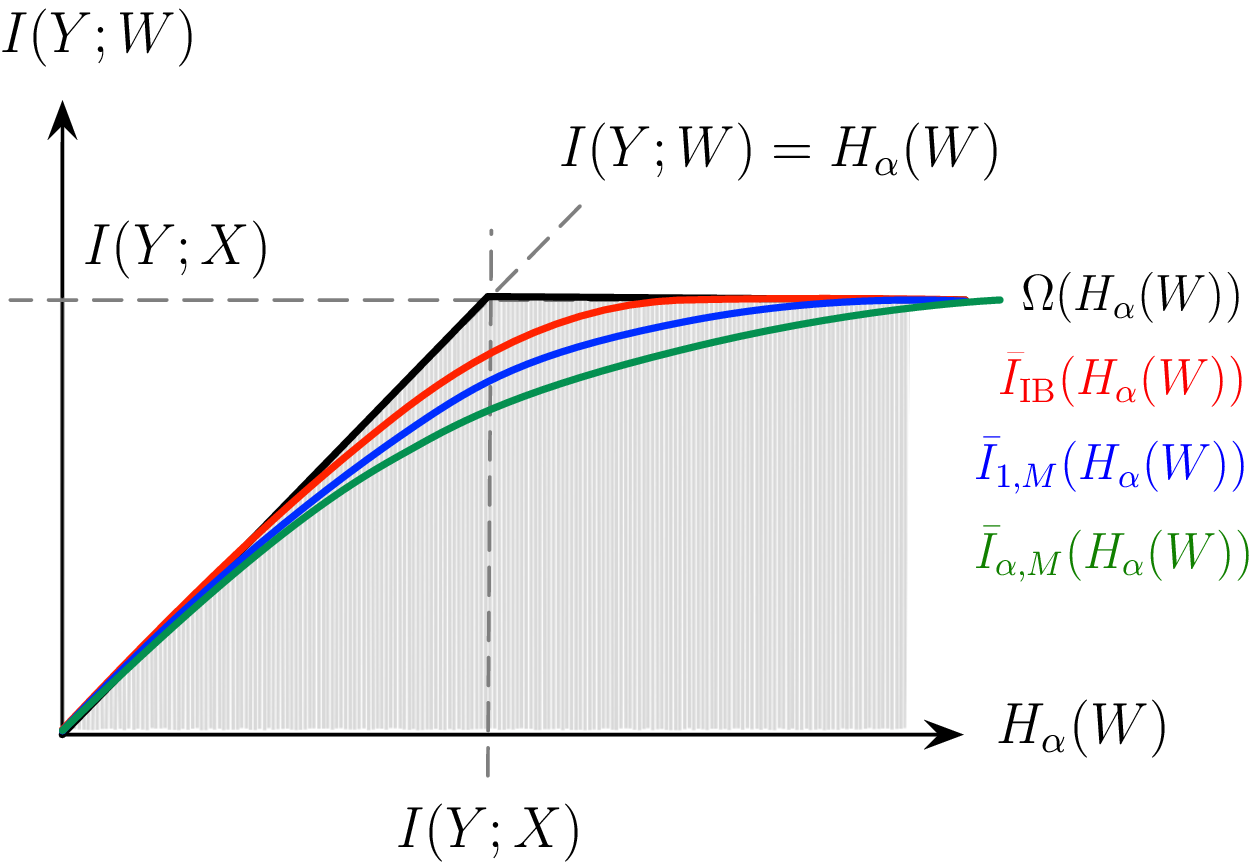}
        \caption{}
        \label{fig:REIBOuter}
     \end{subfigure}
    \hfill\hspace{+0.8cm}
    \begin{subfigure}[b]{0.3\textwidth}
      \centering
      \includegraphics[draft=false, scale=0.435]{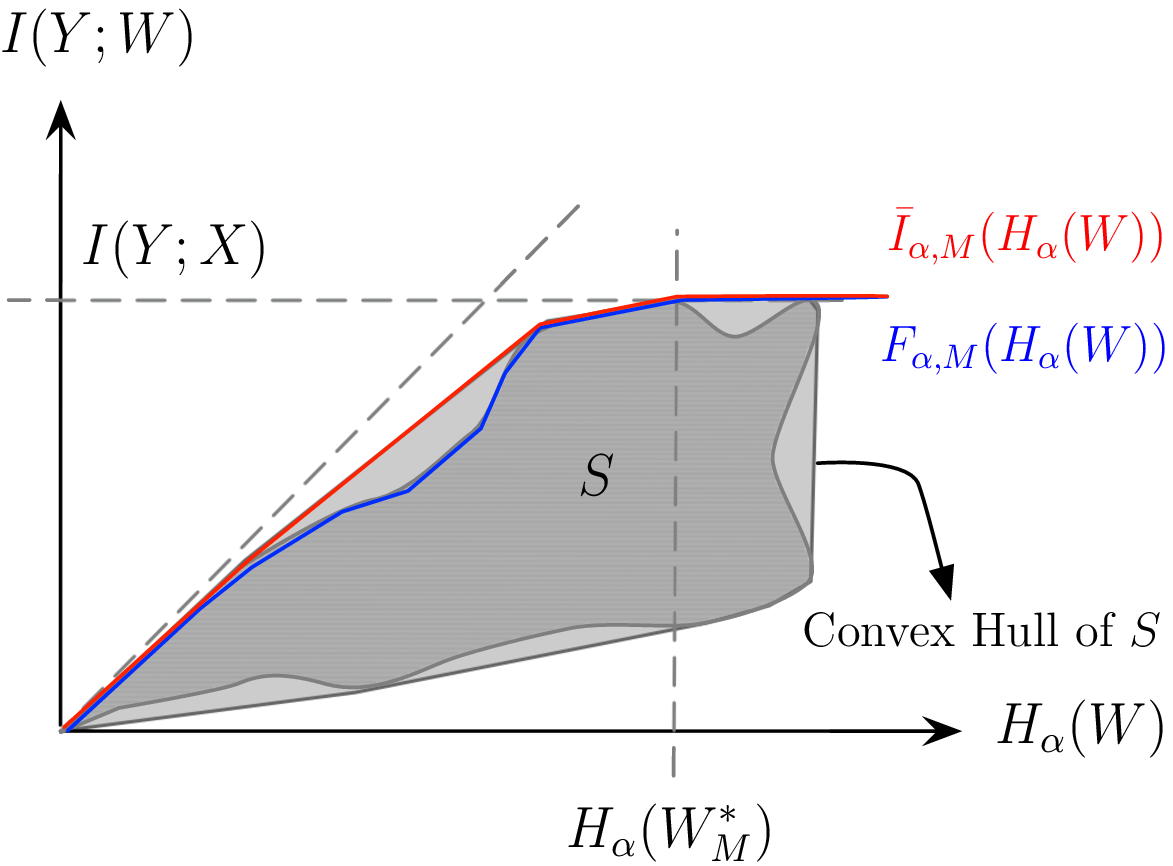}
      \caption{}
      \label{fig:IEIB1}
    \end{subfigure}
    \hfill
    \begin{subfigure}[b]{0.3\textwidth}
      \centering
      \includegraphics[draft=false, scale=0.435]{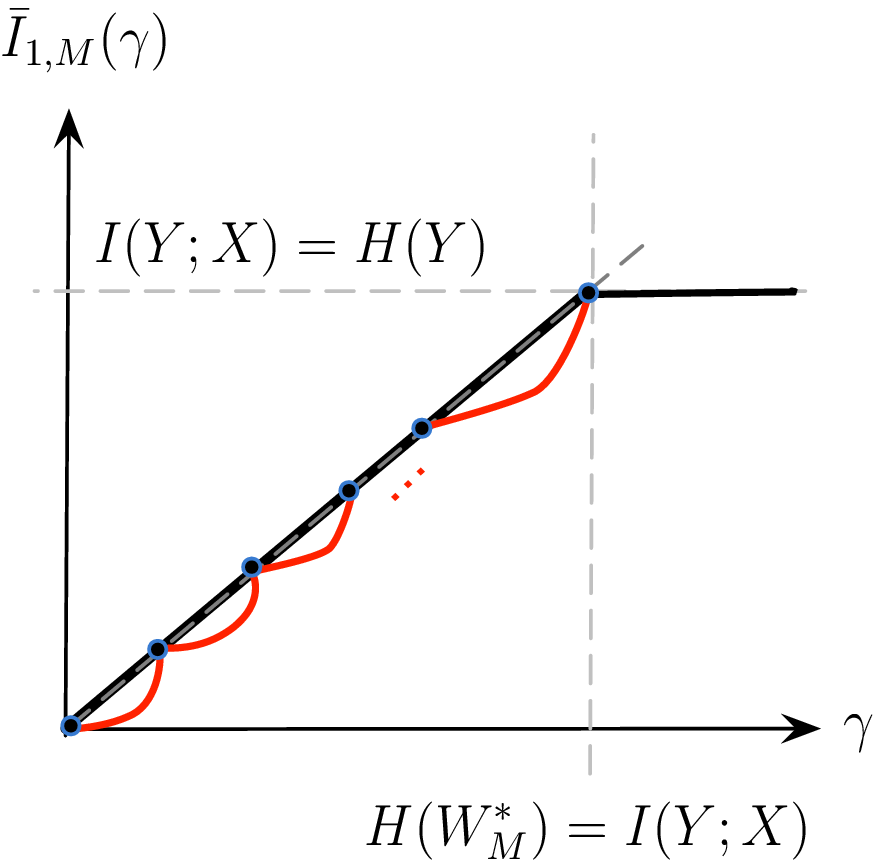}
      \caption{}
      \label{fig:IEIB2}
    \end{subfigure}
  \caption{(a) All feasible R\'{e}nyi entropy-relevance pairs must lie in the shaded region, (b) Typical shapes of the set $S=\{(H_{\alpha}(W), I(Y;W)): P_{W|X}\in\mathcal{P}(\mathcal{W}|\mathcal{X})\}$ and the convex hull of $S$, and (c) The plot of $\bar{I}_{1,M}(\gamma)$ for the case $Y=f(X)$, where the black dots are given by deterministic $P_{W|X}$'s and the curve is obtained by the DIB method \cite{kolchinsky2018}.}\vspace{-0.4cm}
  \label{fig:EIB}
\end{figure*}

\section{The R\'{e}nyi Entropy-Relevance Trade-off and Operational Characterization}
In Section~I, we defined $\bar{I}_{\alpha, M}$ as the upper concave function of $F_{\alpha, M}$. 
In this section, we investigate the properties of $\bar{I}_{\alpha, M}$, which will be used in deriving an operational characterization and to develop efficient numerical methods to estimate $\bar{I}_{\alpha, M}$. 
Below, we assume that $P_{Y, X}=P_{X}P_{Y|X}$, $M<\infty $, and $\alpha\in(0, 1)$ are fixed, unless otherwise stated. 

\subsection{Bounds on $\bar{I}_{\alpha, M}(\gamma)$ and Properties of $\bar{I}_{\alpha, M}(\gamma)$}
We first recall under the Markov chain constraint $Y\markov X\markov W$, we have 
\begin{IEEEeqnarray}{rCl}
  \subnumberinglabel{eq:bounds}
  I(Y; W) &\le & I(Y; X)\label{eq:bound1}
  \vspace{-0.2cm}
\end{IEEEeqnarray}
and\vspace{-0.2cm}
\begin{IEEEeqnarray}{rCl}
  \IEEEyessubnumber*
  I(Y; W) &\le & H(W) \le H_{\alpha}(W)\le \log_2 |\mathcal{W}|\label{eq:bound2}
\end{IEEEeqnarray} 
where \eqref{eq:bound1} is due to the data-processing inequality \cite{cover2006} and \eqref{eq:bound2} holds since $H_{\alpha}(W)$ is non-increasing in $\alpha$. 
Based on \eqref{eq:bounds}, we construct a concave function $\Omega$ given by $\Omega(\gamma)=\gamma$ for $\gamma\in[0, I(Y; X)]$ and $\Omega(\gamma)=I(Y; X)$ for $\gamma> I(Y; X)$. Using the definition of $\bar{I}_{\alpha, M}$ and the concavity of $\Omega$, we immediately obtain the following result. 

\begin{lemma}
  $\bar{I}_{\alpha, M}(\gamma)\le \Omega(\gamma)\le \log_2 M$ for any $\gamma\ge 0$. 
  \label{lma:ub1}
\end{lemma}

Moreover, we can relate $\bar{I}_{1, M}$ to the IB function $\bar{I}_{\text{IB}}$ in \eqref{eq:IB}: 
\begin{lemma}
  $\bar{I}_{1, M}(\gamma)\le \bar{I}_{\emph{IB}}(\gamma)$ for any $\gamma\ge 0$. 
  \label{lma:ub2}
\end{lemma}
\begin{IEEEproof}
  For any $\gamma\ge 0$,  since the constraint $H(W)\le \gamma$ implies that $I(X; W)\le \gamma$, we have that $\bar{I}_{\text{IB}}(\gamma)\ge F_{\alpha, M}(\gamma)$. By invoking the concavity of $\bar{I}_{\text{IB}}$ and the definition of $\bar{I}_{1, M}$, we immediately obtain the desired inequality.    
\end{IEEEproof}

Using a similar argument, one can also deduce the following two statements whose proofs we omit for simplicity. 

\begin{lemma}
  $\bar{I}_{\alpha, M}(\gamma)\le \bar{I}_{1, M}(\gamma)$ for any $\gamma\ge 0$.
  \label{lma:ub3}
\end{lemma}

\begin{lemma}
   $\bar{I}_{\alpha, M}(\gamma)\le\bar{I}_{\alpha, M+1}(\gamma)$ for any $\gamma\ge 0$.
   \label{lma:ub4}
\end{lemma}

A visualization of these bounds is given in Fig.\!~\ref{fig:REIBOuter}. 
In the following, we present some properties of $\bar{I}_{\alpha, M}$.
First, $\bar{I}_{\alpha, M}(\gamma)$ is non-decreasing in $\gamma$ since $F_{\alpha, M}(\gamma)$ is non-decreasing. 
Moreover, $\bar{I}_{\alpha, M}(\gamma)$ is a continuous function in $\gamma$. To see this, we note that $\bar{I}_{\alpha, M}(\gamma)$ is continuous for $\gamma>0$ due to the concavity \cite[Theorem 10.2]{rockafellar1970}. One can directly verify that $\lim_{\gamma\to 0^+}\bar{I}_{\alpha, M}(\gamma)=\bar{I}_{\alpha, M}(0)$ to complete the proof. 

Furthermore, let $\mathcal{P}_{M}(\mathcal{W}|\mathcal{X})\:=\{P_{W|X}\!:\!|\mathcal{W}| = M\}$. 
Due to the compactness of $\mathcal{P}_M(\mathcal{W}|\mathcal{X})$, the image $S$ of $\mathcal{P}_M(\mathcal{W}|\mathcal{X})$ under the continuous mapping $P_{W|X}\mapsto(H_{\alpha}(W), \allowbreak I(Y; W))$ is also compact, which guarantees the existence of a $P_{W^*_M|X}\in\mathcal{P}_M(\mathcal{W}|\mathcal{X})$ that satisfies $I(Y; W^*_M)=\max(\eta: (\gamma, \eta)\in S):=J$ and 
$H_{\alpha}(W^*_M)=\min(\gamma: (\gamma, J)\in S)$.  
This result indicates that the graph of $\bar{I}_{\alpha, M}$ is flat with $\bar{I}_{\alpha, M}(\gamma)\! =\!J$ for $\gamma\!\ge\! \allowbreak H_{\alpha}(W^*_M)$. 
When $M\ge|\mathcal{X}|$, we have that $J=I(Y; X)$; a visualization of this case is given in Fig.~\ref{fig:IEIB1}. Note that the $W^*_M$ does not necessarily equal to $X$ (in distribution). 


We next establish an operational characterization of $\bar{I}_{\alpha, M}$ by associating each pair $(\gamma, \bar{I}_{\alpha, M}(\gamma))$ with an optimal operational scheme. This result implies that $\bar{I}_{\alpha, M}(\gamma)$ is the maximal achievable relevance of a system when the representation cost is at most $\gamma$ (bits/input symbol). 


\subsection{An Operational Characterization of $\bar{I}_{\alpha, M}(\gamma)$}\label{sec:EIBCodingThm}
We consider the system model in Section~II. 
The goal of an operational scheme here is to transform $X^n$ into $W^n$ under a R\'{e}nyi entropy constraint where $\alpha\in(0, 1)$ while preserving the relevant information about $Y^n$ as much as possible. 
For this purpose, we define R\'{e}nyi entropy-relevance codes below.  

\begin{definition}
  A length-$n$ R\'{e}nyi entropy-relevance code is a mapping $\Phi^n_{\alpha, M}: \mathcal{X}^n\to\mathcal{W}^n$.
\end{definition}
Let $W_i$ denote the $i$th component of the output $\Phi^n_{\alpha, M}(X^n)$. 
The average output R\'{e}nyi entropy $\gamma_n$ associated with $\Phi^n_{\alpha, M}$ is then given by $\gamma_n=\frac{1}{n}\sum_{i=1}^n H_{\alpha}(W_i)$, and the associated average relevance level $\eta_n$ is computed as $\eta_n=\frac{1}{n}\sum_{i=1}^{n} I(Y_i; W_i)$. 

\begin{definition}
  A R\'{e}nyi entropy-relevance pair $(\gamma, \eta)$ is said to be achievable if there exists a sequence of codes $\{\Phi^n_{\alpha, M}\}$ such that $\limsup_{n\to\infty}\gamma_n\le \gamma$ and $\liminf_{n\to\infty}\eta_n\ge \eta$. The achievable R\'{e}nyi entropy-relevance region $\mathcal{R}_{\alpha,M}\subseteq\mathbb{R}^2_{\ge 0}$ is defined as the closure of all achievable pairs.   
\end{definition}

Similar to the IB result, we next define $I_{\alpha,M}(\gamma)=\max(\eta: (\gamma, \eta)\in\mathcal{R}_{\alpha,M})$. 
We obtain the following theorem. 
\begin{theorem} 
  $I_{\alpha, M}(\gamma)=\bar{I}_{\alpha, M}(\gamma)$ for $\gamma\ge 0$.  
  \label{thm:EIBCodingThm}
\end{theorem}
\begin{IEEEproof}
(\emph{Achievability}): It suffices to consider the situation where $\gamma\le H_{\alpha}(W^*_M)$ since by the flatness property of the function $\bar{I}_{\alpha, M}$ and Definition~4, the achievability of the pair $(H_{\alpha}(W^*_{M}), \bar{I}_{\alpha, M}(H_{\alpha}(W^*_M)))$ implies the achievability of $(\gamma, \bar{I}_{\alpha,M}(\gamma))$ for any $\gamma> H_{\alpha}(W^*_M)$.  
Now we show that any pair $(\gamma, \eta)=(\gamma, \bar{I}_{\alpha, M}(\gamma))$ with $\gamma\le H_{\alpha}(W^*_M)$ is achievable. 
By the definition of $\bar{I}_{\alpha, M}(\gamma)$ in \eqref{eq:barI}, we can write $(\gamma, \bar{I}_{\alpha, M}(\gamma))=\sum_{k=1}^2 \lambda_k(\gamma^{(k)}, F_{\alpha, M}(\gamma^{(k)}))$ for some pair $(\gamma^{(k)}, F_{\alpha, M}(\gamma^{(k)}))$,  $\lambda_k\ge 0$, $k=1, 2$, and $\lambda_1+\lambda_2=1$. 
Suppose that the conditional probability $P_{W^{(k)}|X}$ determines $(\gamma^{(k)}, F_{\alpha, M}(\gamma^{(k)}))$. Note that such $P_{W^{(k)}|X}$ exists due to the definition of $F_{\alpha, M}$.

We next construct a code $\Phi^n_{\alpha, M}$ using time-sharing \cite{cover2006}.
Specifically, the input sequence $X^n$ is divided into two disjoint sub-blocks and the size of the $k$th sub-block is $n\alpha_k$. In the $k$th sub-block, $k=1, 2$, our code $\Phi^n_{\alpha, M}$ maps each $X_i$ into $W_i$ symbol-wise according to $P_{W^{(k)}|X}$. 
Clearly, using this coding scheme, the average output R\'{e}nyi  and the average relevance of the $k$th sub-block will be $\gamma^{(k)}$ and $\eta^{(k)}$, respectively, implying that the pair $(\gamma, \eta)$ is achievable.\smallskip

\noindent(\emph{Converse}):
We claim that any achievable pair $(\gamma, \eta)$ satisfies $\eta\le \bar{I}_{\alpha,M}(\gamma)$. 
Given $(\gamma, \eta)$-achievable codes $\{\Phi^n_{\alpha, M}\}$, we proceed the following standard steps to prove the claim: 
\begin{IEEEeqnarray}{l}
    \eta_n=\frac{1}{n}\sum_{i=1}^n I(Y_i; W_i)\le  \frac{1}{n}\sum_{i=1}^n \bar{I}_{\alpha, M}(H_{\alpha}(W_i))\nonumber\\
    \qquad\quad\ \ \le  \bar{I}_{\alpha,M}\scalemath{1}{\Bigg(\frac{1}{n}\sum_{i=1}^n H_{\alpha}(W_i)\Bigg)}= \bar{I}_{\alpha, M}(\gamma_n),\nonumber
\end{IEEEeqnarray}
  where the second inequality holds since $\bar{I}_{\alpha,M}$ is concave while others hold by definition. 
  The claim is proved by noting that
\begin{IEEEeqnarray}{l}
  \eta\le \liminf_{n\to\infty} \eta_n \le \liminf_{n\to\infty} \bar{I}_{\alpha, M}(\gamma_n)= \bar{I}_{\alpha,M}\bigg(\liminf_{n\to\infty} \gamma_n\bigg)\nonumber\IEEEeqnarraynumspace\\
  \qquad\quad\le\bar{I}_{\alpha, M}\bigg(\limsup_{n\to\infty} \gamma_n\bigg)\le \bar{I}_{\alpha,M}(\gamma),\nonumber\IEEEeqnarraynumspace
\end{IEEEeqnarray}
where we have used the continuity and non-decreasing properties of $\bar{I}_{\alpha,M}$ and the definition of achievability. 
\end{IEEEproof}

\section{Numerical Methods and Examples} 
Due to an entropy mismatch for $W$ in the objective function $I(Y;W)=\allowbreak H(W)-\allowbreak H(W|Y)$ and the constraint $H_{\alpha}(W)\le\gamma$ of \eqref{eq:GEIBC}, an analytical expression for $\bar{I}_{\alpha, M}$ is difficult to derive.  
This section discusses numerical methods for approximating $\bar{I}_{\alpha, M}$. Two special examples where $\bar{I}_{1,M}$ can be exactly determined are given and some numerical results are also presented. 
To approximate $\bar{I}_{\alpha, M}$, we consider the maximization problem
\begin{IEEEeqnarray}{C}
  \max_{\substack{P_{W|X}: Y\markov X\markov W\\ |\mathcal{W}|=M}} \left[\beta I(Y; W)- H_{\alpha}(W)\right]
  \label{eq:maxi}
\end{IEEEeqnarray}
where $\beta\in[0, \infty)$ controls the trade-off between $I(Y; W)$ and $H_{\alpha}(W)$. 
For a fixed $\beta$, the maximizer $P_{W^*|X}$ of \eqref{eq:maxi} will result in a R\'{e}nyi entropy-relevance pair $(H_{\alpha}(W^*), I(Y;W^*))$.\footnote{We note that this pair $(H_{\alpha}(W^*), I(Y;W^*))$ is an extreme point of the hypograph of $\bar{I}_{\alpha, M}(\gamma)$ picked out by a support line of slope $\frac{1}{\beta}$.} 
When $\beta=0$, one obtains the trivial pair $(0, 0)$.
Moreover, the argument in Section~\ref{sec:IIB} implies that the maximizer of \eqref{eq:maxi} is deterministic. Thus, one can estimate $\bar{I}_{\alpha, M}(\gamma)$ by varying $\beta$ and checking $|\mathcal{W}|^{|\mathcal{X}|}$ possible deterministic mappings for each fixed $\beta$. 
Specifically, let $S'$ denote the set of all obtained R\'{e}nyi entropy-relevance pairs. 
For each $\gamma\ge 0$, the estimation of $\bar{I}_{\alpha, M}(\gamma)$ is then given by $\max(\eta:(\gamma', \eta)\in S', \gamma'\le \gamma)$.   
Clearly, the overall procedure is quite complex. 

Next we provide an iterative algorithm that avoids checking all $|\mathcal{W}|^{|\mathcal{X}|}$ possible mappings for a fixed $\beta$. The algorithm is derived using the first-order optimality condition \cite{boyd2004} for the following modified Lagrangian of \eqref{eq:maxi}: 
\begin{IEEEeqnarray}{l}
  \mathcal{L}(\nu, \bm{\mu}, \beta, \alpha, P_{W|X})=\beta I(Y;W)-\nu H(W|X)\qquad\nonumber\IEEEeqnarraynumspace\\
  \qquad\qquad\qquad- H_{\alpha}(W)-\sum_{x\in\mathcal{X}}\mu(x)\sum_{w\in\mathcal{W}}P_{W|X}(w|x)\IEEEeqnarraynumspace
  \label{eq:Lfunc}
\end{IEEEeqnarray}
where $\nu\ge 0$ and $\bm{\mu}$ is a vector containing the Lagrangian multipliers $\mu(x)$, $x\in\mathcal{X}$, for the constraint $\sum_{w}P_{W|X}(w|x)=1$. As $\nu\to 0$, the function $\mathcal{L}$ converges to the Lagrangian of \eqref{eq:maxi}. 
Note that the term $-\nu H(W|X)$ is added to obtain an explicit expression of $P_{W|X}$ in \eqref{eq:sol} below. Setting $\pdv{\mathcal{L}}{P_{W|X}(w|x)}=0$ for each pair $w$ and $x$, we obtain the following consistency equation for the maximizer $P_{W|X}$ of $\mathcal{L}$:
\begin{IEEEeqnarray}{l}
  P_{W|X}(w|x)=\scalemath{0.95}{\frac{1}{Z}\exp\bigg[\frac{-1}{\nu}\bigg(\frac{\alpha (P_{W}(w))^{\alpha-1}}{(1-\alpha)\sum_{w\in\mathcal{W}}(P_{W}(w))^{\alpha}}}\nonumber\\
  \qquad\qquad\qquad\qquad\qquad\qquad +\scalemath{0.95}{\beta D(P_{Y|X=x}||P_{Y|W=w})\bigg)\bigg]}\IEEEeqnarraynumspace
  \label{eq:sol}
\end{IEEEeqnarray}
where $Z:=Z(x, \nu, \bm{\mu}, \alpha)$ is a normalization factor and $D(\cdot||\cdot)$ denotes the Kullback–Leibler divergence \cite{cover2006}. 
Letting $\nu\to 0$ in \eqref{eq:sol} for all $w$ and fixed $x$, one easily observes that the optimal $P_{W|X}$ tends to be deterministic, which coincides with the result obtained by applying the convex optimization argument in Section~II-B to \eqref{eq:maxi}. 
Using this observation and \eqref{eq:sol}, we propose the following iterative algorithm to solve \eqref{eq:maxi}. Note that \cite[Algorithm~2]{strouse2017} can be employed to estimate $\bar{I}_{1, M}$. 
\smallskip

\noindent\underline{Initialization}: Randomly generate $P^{(0)}_{W|X}$ and obtain $P^{(0)}_W$ and $P^{(0)}_{Y|W}$ from the probability distribution $P^{(0)}_{W|X}P_{Y, X}$. Set $l=1$.\vspace{+0.1cm} 

\noindent\underline{Step 1}: For each $x\in\mathcal{X}$, compute 
\begin{IEEEeqnarray}{l}  
  w^*(x)=\scalemath{0.95}{\argmax_{w\in\mathcal{W}} \bigg[\frac{\alpha (P^{(l-1)}_{W}(w))^{\alpha-1}}{(1-\alpha)\sum_{w\in\mathcal{W}}(P^{(l-1)}_{W}(w))^{\alpha}}}\nonumber\\
  \qquad\qquad\qquad\qquad\qquad\qquad\quad +\scalemath{0.95}{\beta D(P_{Y|X=x}||P^{(l-1)}_{Y|W=w})\bigg]}\nonumber\vspace*{-0.1cm}%
\end{IEEEeqnarray}
and set $P^{(l)}_{W|X}(w|x)=1$ for $w=w^*(x)$ and set $P^{(l)}_{W|X}(w|x)=0$ for all $w\neq w^*(x)$.\vspace{+0.1cm}

\noindent\underline{Step 2}: If $P^{(l)}_{W|X}=P^{(l-1)}_{W|X}$ or the maximum number of iterations is reached, then terminate the procedure and compute the pair $(H_{\alpha}(W), I(Y;W))$ using $P^{(l)}_{W|X}P_{Y, X}$. Otherwise, obtain the marginal probability distributions $P^{(l)}_{W}$ and $P^{(l)}_{Y|W}$ from $P^{(l)}_{W|X}P_{Y, X}$, set $l=l+1$, and go back to Step 1. \vspace{+0.15cm}

We remark that the above iterative algorithm may only yield a local maximizer for \eqref{eq:maxi}. To alleviate this situation, one can initialize this algorithm with different $P^{(0)}_{W|X}$ and choose the best result. 
Next, we determine $\bar{I}_{1, M}$ for two special cases. 

\begin{example}
  Given a function $f$, consider $Y=f(X)$. Then, we have that $\bar{I}_{1,M}(\gamma)=\Omega(\gamma)$ 
    for any finite $M\ge|\mathcal{X}|$; the function $\bar{I}_{1,M}$ is drawn in Fig.~\ref{fig:IEIB2}. Based on \eqref{eq:bounds}, it suffices to show that $\bar{I}_{1, M}(0)=0$ and $\bar{I}_{1, M}(I(Y;X))=I(Y;X)$, where $I(Y;X)=H(Y)$.   
  The former case is apparent since $H(W)=0$ implies that $I(Y; W)=0$ and hence $\bar{I}_{M}(0)=0$. 
  For the latter case, we set $W=h(f(X))$ for some injective function $h:\mathcal{Y}\to\mathcal{W}$. 
  The desired result simply follows from the fact that the $P_{W|X}$ induced by the  mapping $W=h(f(X))$ achieves the upper bound in \eqref{eq:bound1} since   
  \begin{IEEEeqnarray}{rCl}
    H(W)&=& H(h(f(X)))=H(h(Y))=H(Y),\nonumber\\
    I(Y; W)&=& H(Y)-H(Y|W)=H(Y),\nonumber
  \end{IEEEeqnarray}
  where $H(Y|W)=H(Y|h(Y))=0$ since $h$ is injective. 
\end{example}

When comparing our result with the DIB result in Fig.~\ref{fig:IEIB2}, we observe that $\bar{I}_{1, M}$ attains a higher relevance level given the same Shannon entropy constraint. 
In fact, this increment of relevance level mainly comes from the convexification of the DIB curve. Such an operation is missing in the context of the DIB method due to the lack of an operational interpretation. Here, our Theorem~\ref{thm:EIBCodingThm} provides the rationale for this operation and indicates that one can use time-sharing between two DIB schemes to take a better Shannon entropy-relevance trade-off. 

\begin{example}
  Suppose that the given joint probability matrix $[P_{Y, X}(\,\cdot\,, \,\cdot\,)]$ can be arranged into a diagonal form as shown in Table~\ref{tab:1a} with the maximum possible number $K$ of non-zero blocks and the $k$th block contains identical probability mass $p_k$, $1\le k\le K$. 
  Choose $f_1:\mathcal{X}\to\allowbreak\{1, 2, \dots, K\}$ and $f_2:\mathcal{Y}\to\allowbreak\{1, 2, \dots, K\}$ such that $P_{f_1(X)}$ and $P_{f_2(Y)}$ are strictly positive and $f_1(x)=f_2(y)$ whenever $P_{X, Y}(x, y)>0$ for all $x$ and $y$. 
  A choice of such $f_1$ and $f_2$ is given in Table~\ref{tab:1b}.
  Moreover, let $\mathcal{X}_k\! =\!\{x\!\in\!\mathcal{X}\!:\!f_1(x)\!=\!k\}$ and $\mathcal{Y}_k\! =\!\{y\!\in\!\mathcal{Y}\!:\! f_2(y)\! =\!k\}$, and set $s_k=\allowbreak\Pr(X\in\mathcal{X}_k, Y\in\mathcal{Y}_k)=|\mathcal{X}_k||\mathcal{Y}_k|p_k$. 
  We can then show that $\bar{I}_{1, M}(\gamma)=\Omega(\gamma)$
  for any finite $M\ge K$. 
  The details are provided in the Appendix.
\end{example}

To end this section, we apply our iterative algorithm to estimate $\bar{I}_{\alpha, M}$ for the $P_{Y, X}$ in Table~I(a) with $\alpha\in\{0.1, 0.5\}$ and $M\in\{2, 3\}$. Here, $H(X)=2.25$ bits and $I(Y; X)=1.5$ bits. Our estimation for the different $\bar{I}_{\alpha, M}$'s are depicted in Fig.~3. When $M=2$ and $\beta\ge 1$, our algorithm produces a maximizer $P_{W|X}$ that induces a uniform distribution on $\mathcal{W}$, regardless of the value of $\alpha$. This maximizer corresponds to the R\'{e}nyi entropy-relevance pair $(1, 1)$. Together with the trivial pair $(0,0)$, we obtain an estimate for $\bar{I}_{\alpha, 2}$. In fact, Lemmas~\ref{lma:ub1} and~\ref{lma:ub3} imply that our estimation is exact in this case. Next, when considering $M=3$, we observe that  maximum relevance is attained. The required $H_{\alpha}(W^*_M)$ varies with $\alpha$, but all of them are less than $H(X)$. This result shows that our bottleneck method can also effectively extract relevance information while minimizing the representation cost. 


\section{Conclusion}
Unlike the IB method that characterizes the optimal complexity-relevance  trade-off via a single optimization problem, our bottleneck problem needs additional convexification to describe the optimal R\'{e}nyi entropy-relevance trade-off. 
Our optimal operational scheme consists of two symbol-wise transformations that operate in a time-sharing manner. 
Though not discussed here, our information-R\'{e}nyi entropy method has been applied to geometric clustering and shows a robustness result when the probability distributions of the data model and the data sets are mismatched \cite{adan2018}.  
Still, it remains unclear how the optimal R\'{e}nyi entropy-relevance trade-off vary with $\alpha$ and $M$, which we leave for future research. 
Other research topics include extending our approach to the variational IB problem \cite{alemi2016} and applying our result to other tasks in machine learning.

\begin{table}[t!]
  \quad\scalebox{0.9}{
  \begin{subtable}[b]{0.19\textwidth}
      \centering
      \begin{tabular}{c|ccccc}
        $P_{Y, X}$ & $1$                                & $2$           & $3$                                & $4$           & $5$                                \\ \hline
        $1$        & \multicolumn{1}{l|}{$\frac{1}{4}$} & $0$           & $0$                                & $0$           & $0$                                \\ \cline{2-4}
        $2$        & \multicolumn{1}{l|}{$0$}           & $\frac{1}{8}$ & \multicolumn{1}{l|}{$\frac{1}{8}$} & $0$           & $0$                                \\
        $3$        & \multicolumn{1}{l|}{$0$}           & $\frac{1}{8}$ & \multicolumn{1}{l|}{$\frac{1}{8}$} & $0$           & $0$                                \\ \cline{3-6} 
        $4$        & $0$                                & $0$           & \multicolumn{1}{l|}{$0$}           & $\frac{1}{8}$ & \multicolumn{1}{l|}{$\frac{1}{8}$} \\ \cline{5-6} 
        \end{tabular}
     \caption{$P_{Y, X}$}
     \label{tab:1a}
  \end{subtable}}
  \qquad\quad\ 
  \scalebox{0.9}{
  \begin{subtable}[b]{0.19\textwidth}
      \centering
        \begin{tabular}{cccccc}
          &     &     &     &     &     \\
        \multicolumn{1}{c|}{$X$}      & $1$ & $2$ & $3$ & $4$ & $5$ \\ \hline
        \multicolumn{1}{c|}{$f_1(X)$} & $1$ & $2$ & $2$ & $3$ & $3$ \\
                                      &     &     &     &     &     \\
        \multicolumn{1}{c|}{$Y$}      & $1$ & $2$ & $3$ & $4$ &     \\ \cline{1-5}
        \multicolumn{1}{c|}{$f_2(Y)$} & $1$ & $2$ & $2$ & $3$ &    
        \end{tabular}
      \caption{$f_1$ and $f_2$}
      \label{tab:1b}
   \end{subtable}}
   \caption{An illustration of Example~2. Here, $K=3$, $p_1=\frac{1}{4}$, $p_2=p_3=\frac{1}{8}$, $s_1=\frac{1}{4}$, $s_2=\frac{1}{2}$, and $s_3=\frac{1}{4}$.}
   \label{tab:ex2}
\end{table}
\begin{figure}
  \centering
  \vspace*{-0.4cm}\includegraphics[scale=0.41, draft=false]{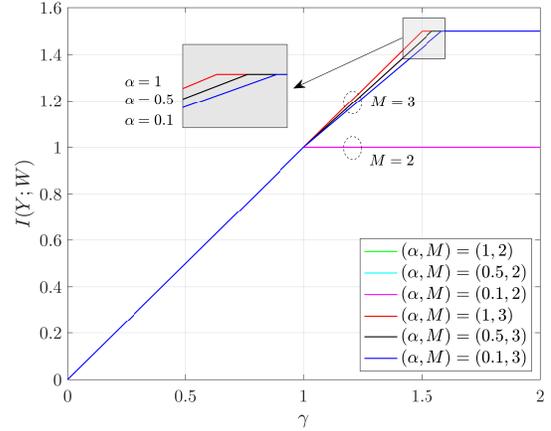}
  \caption{The estimated $\bar{I}_{\alpha, M}$ for $P_{Y, X}$ given in Table~1(a), where $M=\{2,3\}$ and $\alpha=\{0.1, 0.5, 1\}$. Our estimation for $\bar{I}_{\alpha, 2}$ are tight, and the curves are overlapping. Our estimate of $\bar{I}_{\alpha, M}$ for $M\ge 3$ are identical for each given $\alpha$ for this $P_{Y, X}$.}\vspace{-0.45cm}
\end{figure}

\begin{appendix}[Supplementary Result for Example 2]
Similar to Example~1, it suffices to show that $\bar{I}_{1, M}(0)=0$ and $\bar{I}_{1, M}(I(Y; X))\allowbreak=I(Y; X)$. The former case is clear and thus omitted, but for the latter case, we need the following result.  
Given any injective function $h:\{1, 2, \dots, K\}\to\mathcal{W}$, the function $g:\mathcal{X}\to\mathcal{W}$ defined as $g(x)=h(f_1(x))$ yields:   
\begin{IEEEeqnarray}{l}
  I(Y;W)=H(W)-H(W|Y)\nonumber\\
  \ \ \ =H(W)-H(h(f_1(X))|Y)\nonumber\\
  \ \ \ =H(W)-H(h(f_2(Y))|Y)=H(W).\nonumber
\end{IEEEeqnarray}

Without loss of generality, choose $g(x)=k$ for $x\in\mathcal{X}_k$. 
We prove explicitly that $H(W)=I(X; Y)$. 
First,   
\begin{IEEEeqnarray}{l}
  I(Y; X)= H(Y)-H(Y|X)\nonumber\\
  \ = -\sum_{k=1}^K \underbrace{|\mathcal{Y}_k|(|\mathcal{X}_k|p_k)}_{=s_k}\log_2(|\mathcal{X}_k|p_k)-\sum_{k=1}^K\underbrace{|\mathcal{X}_k||\mathcal{Y}_k|p_k}_{=s_k}\log_2|\mathcal{Y}_k|\nonumber\\
  \ = -\sum_{k=1}^K s_k\log_2 s_k.\nonumber
\end{IEEEeqnarray}
Moreover, since $w=g(x)=k$ for $x\in\mathcal{X}_k$, we obtain that 
\[P_{W}(k) =\sum_{y\in\mathcal{Y}}\sum_{x\in\mathcal{X}}P_{Y, X}(y, x)P_{W|X}(k|x)=|\mathcal{Y}_k||\mathcal{X}_k|p_k=s_k.\]
and hence $I(Y; X)=H(W)=I(Y; W)$. Based on the bound in \eqref{eq:bound2}, the equality $\bar{I}_{1, M}(I(Y; X))=I(Y; X)$ clearly holds. 
\end{appendix}

\newpage
\balance
\bibliographystyle{IEEEtran}
\bibliography{../literatureDB}
\end{document}